\documentclass[graybox]{svmult}

\usepackage{type1cm}        %
\usepackage{makeidx}         %
\usepackage{graphicx}        %
\usepackage{multicol}        %
\usepackage[bottom]{footmisc}%

\makeindex             %

\begin{document}

\title*{Urban Scaling Laws}
\author{Fabiano L. Ribeiro and Vinicius M. Netto}
\institute{Fabiano L. Ribeiro \at Department of Physics, Universidade Federal de Lavras (UFLA), Brazil. 
\email{fribeiro@ufla.br}
\and Vinicius M. Netto \at Research Centre for Territory, Transports and Environment (CITTA), Faculty of Engineering (FEUP), University of Porto, Portugal. \email{vmnetto@fe.up.pt}}
\maketitle

\abstract{
Understanding how size influences the internal characteristics of a system is a crucial concern across various fields. Concepts like scale invariance, universalities, and fractals are fundamental to this inquiry and find application in biology, physics, and particularly urbanism. Size profoundly impacts how cities develop and function economically and socially. For example, what are the pros and cons of residing in larger cities? Is life really more expensive or less safe in larger cities? Or do they really offer more opportunities and generally higher incomes than smaller ones? To address such inquiries, we utilize theoretical tools from scaling theory, enabling a quantitative description of how a system's behavior changes across different scales, from micro to macro. Drawing parallels with research in biology and spatial economics, this chapter explores recent discoveries, ongoing progress, and unanswered questions regarding urban scaling.
}

\section{Introduction}
\label{sec_intro}
Understanding how size affects the internal properties of a system is a critical issue in many areas of knowledge. For instance, when a new drug is tested in mice, how is it possible to extrapolate the same properties to the human body, which is three orders of magnitude larger? Or, in the context that we are interested in here, what are the advantages or disadvantages of living in a city larger than another? That is, is living in a bigger city more expensive or dangerous, or does it bring more opportunities for interaction and materializing ideas or even achieving a higher income than living in a smaller one? How can we know if a city of a given population needs one more hospital or a petrol station? To answer these questions, we need to use a set of theoretical tools known as \emph{scaling theory} \cite{Kadanoff2000, Stanley1987, West2017}, which allows for a quantitative description of how a system changes its behavior from one scale to another, or, more precisely, from a micro-scale to a macro-scale. Scaling invariance, normalization group, critical phenomena, universalities, and fractals are examples (not exhaustive) of concepts that form the core of this theory, and they find applicability in fields such as biology, physics, engineering, and, for the purpose of this chapter, urbanism.

A naive approach to scaling from one system to another is to perform a simple linear extrapolation of the data. For example, if we effectively use 1\,ml of a drug in a mouse weighing 40\,g, then we might assume that we need to use 1\,L of the same drug for a human weighing 40\,kg (1,000 times bigger). Similarly, if a city of 100,000 inhabitants works well with 10 petrol stations, then we might assume that another city with 1,000,000 inhabitants will work well with 100 petrol stations. However, this simplistic linear extrapolation of the data does not work in these contexts. Biological and urban systems do not scale in a purely multiplicative way, meaning that a simple multiplication of the results obtained by the prototype (a mouse or a small city) will not allow us to obtain the desired effect in another scale of interest (a human body or a large city). In other words, both biological entities and cities are nonlinear scaling systems, and this is what we aim to explore in this chapter.

\begin{figure}[t]
\sidecaption[t]
\includegraphics[scale=0.4]{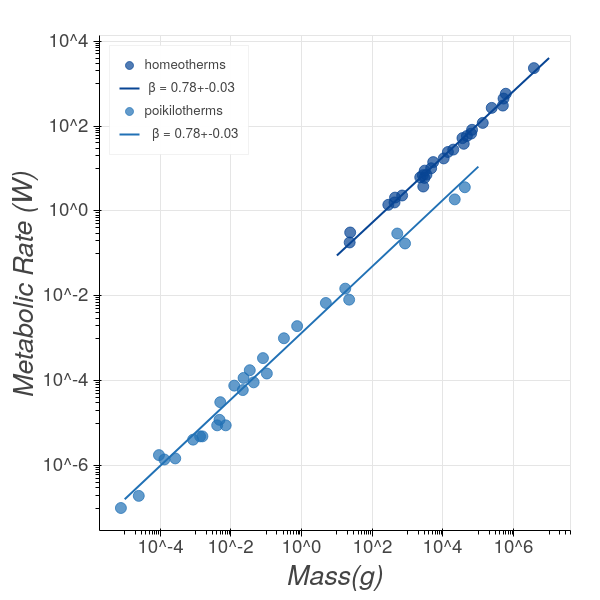}
	\caption{ \label{Fig_lei34} 
Metabolic rate as a function of body mass for two vascular taxonomic groups: homeotherms (organisms that maintain a constant internal body temperature) and poikilotherms (organisms that vary their internal body temperature). A power-law of the type given by Eq.~(\ref{eq_power_law}) is observed for all taxonomic groups (the straight lines capture the trend of points). The parameter $Y_0$ (the intercept) varies from group to group, but $\beta$ is approximately constant and sub-linear ($\beta < 1$). The data were extracted directly from the references \cite{Allman1999} (homeotherms) and \cite{Hemmingsen1960} (homeotherms and poikilotherms). 
Source: \cite[Fig.1]{Ribeiro2022}.}
	\end{figure}

Many systems that present such non-linearity are (usually) governed by a power-law of the form
\begin{equation}\label{eq_power_law}
Y = Y_0 N^{\beta} \,  ,
\end{equation}
where $N$ is the metric that represents the size of the system, $Y$ is a dependent variable, $Y_0$ is a constant, namely the intercept, and $\beta$ is the \emph{scaling exponent}.
In the case of biology, this equation represents well how the metabolic rate, 
which is the amount of energy expended by an organism per time, relates to its body mass \cite{West1997,Ribeiro2022}, as shown in Fig.~\ref{Fig_lei34}. The straight line in a log-log plot is the signature of a system well described by a power-law equation like Eq.~(\ref{eq_power_law}), and this is the case for mass and metabolic rate in biology, as evidenced in this figure. %
For vascular beings larger than $\sim 10^{-4}$\,g, as poikilotherms (animals that vary their internal body temperature) and homeotherms (animals that maintain a constant internal body temperature), we have the so-called \emph{Kleiber's law}, that is the particular case of Eq.~(\ref{eq_power_law})  with $\beta = 3/4$ \cite{shiftspnas2010}.
The fact that $\beta$ is different from 1 (less than 1 in this case) means that the metabolic rate scales non-linearly (sub-linearly in this case) with the organism's size. 
In addition to the metabolic rate, other quantities also scale non-linearly with the size of the individual, such as heart rate and brain size.

\begin{figure}[t]
	\sidecaption[t]
	\includegraphics[scale=0.40]{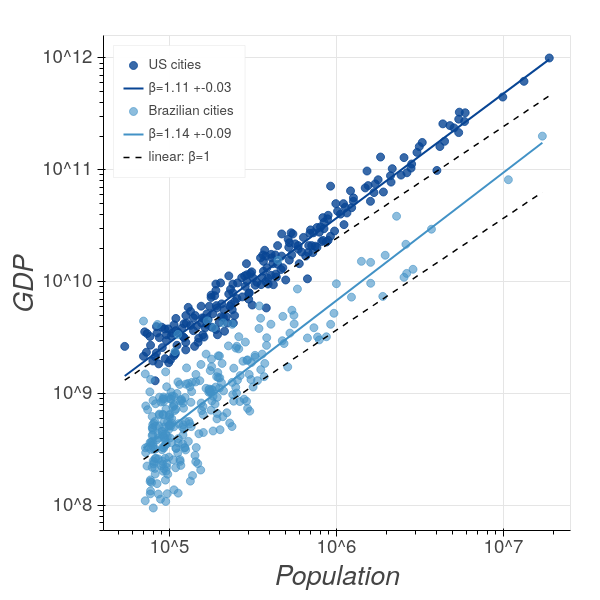}
	\caption{
		\label{fig_pib}
Example of urban scaling.
The urban gross domestic product (GDP) is plotted as a function of population $N$ for cities in Brazil and the United States (on a double logarithmic scale). The solid lines (blue and green) are the best fits that capture the trend of the points and reveal the power-law property. Both countries exhibit scaling exponents $\beta > 1$ (super-linear), with very similar values despite the socio-economic differences between the two countries, which justifies the difference in the vertical position of the straight lines (different intercepts $Y_0$). The dashed lines indicate linearity ($\beta = 1$) and are included only to highlight the super-linear behavior of the data trend. 
Source: \cite[Fig.1]{Ribeiro2023}.
}
\end{figure}

In the case of a city, we have a similar situation when we analyze many urban metrics as a function of the city population size. For instance, Fig.(\ref{fig_pib}) shows that Eq.(\ref{eq_power_law}) describes the empirical data well (straight line in the log-log plot) when $Y$ represents the urban gross domestic product (GDP) and $N$ represents the city population size of cities in Brazil and the United States \cite{bettencourt2007growth, Bettencourt2013, ribeirocity2017}. Despite the fact that the two sets of data are vertically different, characterized by a different intercept due to economic differences between these two countries, the slopes of the straight lines that characterize the data are approximately the same.

These empirical facts suggest that, despite the remarkable complexity underlying biological and urban phenomena, the relationships between metabolic rate and body mass, and between GDP and city population size seem to obey a relatively simple relationship, given by the power-law equation~(\ref{eq_power_law}). 

A power-law equation has some interesting properties, but maybe the most interesting for our purpose here is that it is scale invariant if $\beta$ is fixed.
Let us explain this in more detail. 
Suppose the size metric changes from $N$ to $\lambda N$, where $\lambda$ is an amplification factor. 
Following the equation~(\ref{eq_power_law}), one gets
\begin{equation}\label{Eq_lambda}
   Y(\lambda N) = Y_0 (\lambda N)^{\beta} = Y_0' N^{\beta} 
\end{equation}
where $Y_0' \equiv Y_0 \lambda^\beta$ is another constant. 
Consequently, $Y(\lambda N) \propto Y( N)$ (from~Eq.~(\ref{eq_power_law}) and~(\ref{Eq_lambda})), which means that the behavior of the system does not change when the scale changes. This is the \emph{scale invariance}. 
In other words, one can say that systems which are governed by a power-law, such as biological entities or cities, appear to behave the same way regardless of the scale we use to look at it
\cite{Newman2005}. 
In short, these systems do not have a natural scale and, because of this,  they are also called scale-free systems \cite{barabasi-book}.

To illustrate the concept of scale-invariance, let us present one counterexample and one example where scale-invariance hold true. A person who is 2.5\,m tall (the height of the largest living human being) is undoubtedly an exceptional case, as the distribution of people's heights follows a normal distribution, which has a typical size or scale \cite{Newman2005}.
Therefore, the distribution of people's heights is not scale-invariant. 
However, the distribution of city sizes 
is governed by a power-law equation (the so-called \textit{Zipf's law}, which is explored in more detail in the next section). Consequently,  the city-size distribution is scale-invariant, implying that cities like New York or São Paulo are not exceptional in size. Despite being the largest cities in their respective countries, these cities have the expected sizes given the %
system of cities to which they belong \cite{west-new-york}.

\section{History}
\label{sec_hist}
The attempt to explain how the properties of things change when their size increases is not new. For instance, Galileo studied this question and discussed how the geometry of things limits the existence 
of arbitrarily large living beings \cite{West2017}.
However, it was only in the 19th century that a systematic way to quantify scaling was developed, particularly in biology. 
In fact, a power-law of the form Eq.~(\ref{eq_power_law}) was first observed in biology in 1839 by Saurus and Rameaux \cite{Robiquet1839}. They noticed that the metabolic rate per unit weight decreases with increasing animal size. 
At the end of the 19th century, Max Rubner \cite{Rubner1883,bertalanffy1957} studied dogs and found that the energy production per square meter of the body surface is constant with the animal's size, leading to a power-law relationship between energy and body mass with $\beta = 2/3$. 
Based on this observation, he postulated that living organisms evolved, by natural selection, to a state in which body mass should follow a surface scaling law and thus be able to radiate excess heat. He proposed that the numerical value of the \emph{allometric (scaling) exponent} would be a natural response to the release of heat by the organism, establishing the relationship between the surface area and volume of the organism. This idea led to a theoretical exponent $\beta = 2/3$.
Rubner's theory, known as the \emph{surface hypothesis}, seemed to be reasonably coherent and was accepted for 50~years.
However, more careful experiments were performed at the beginning of the 20th century. Among these studies, we highlight the works of Krogh~\cite{Krogh1916} and Kleiber in 1932 (the best-known study) \cite{Kleiber1932}.
From the analyzed data set, an experimental value of $\beta \approx 3/4$ was observed, which differed from the theoretical result that was accepted until then.
This result is now known as the \emph{$3/4$ Kleiber's law}.
The sub-linear behavior of the scaling exponent ($\beta<1$) implies that larger animals demand less energy per cell, indicating a ``scaling economy''. 
Nowadays, we know that the $3/4$ Kleiber's law applies only to vascular beings and not to prokaryotes and protists organisms, which present super-linear and linear scaling properties, respectively \cite{DeLong2010}.

After an extensive debate on the precise value of the scaling exponent in biology, fundamental work was done in the late 1990s by the theoretical physicist Geoffrey West, together with the biologists James Brown and Brian Enquist. They proposed a model based on the efficiency of nutrient distribution within organisms to explain the scaling law in biology. The theory is based on three premises: (i) a fractal distribution network, (ii) terminal units that do not vary with organism size, and (iii) natural selection, that favors a distribution network that minimizes energy waste. 
This model yields a scaling exponent $\beta = 3/4$ \cite{West1997,West1999,West2004,Savage2008,Ribeiro2022}. However, given its own premises, this theory only applies to organisms with closed circulatory systems.
What is curious about all these ideas is that while Galileo explained the impossibility of arbitrarily large animals using simple geometry, a broader understanding of scaling laws involves more complex concepts, such as networks and fractals.

Continuing with the historical development of scaling understanding,  in 1932
the linguist George Zipf realized when analyzing the frequency of word use in texts that a relatively small set of words are used many times, while the vast majority of words are rarely used \cite{zipf1932}. By quantifying the frequency of word use, he noticed that the most used word in a text, book, or newspaper was, on average, used twice as much as the second most used word and three times as much as the third most used word, and so on. Zipf called this pattern the \emph{Rank versus Frequency rule}, but today it is simply known as \emph{Zipf's law}. 
Zipf (1949) \cite{zipf-book-1949} and (previously) Auerbach (1913) \cite{Auerbach1913,diego2023}
also realized that the same properties are observed in city size distributions. The most populous city in a country is typically twice as large as the second most populous city, and so on, analogously to the frequency of words.
Today, we know that these ranking rules apply to many other contexts, and because of this, power-law distribution with exponent $-1$ are often called \emph{Zipf's law} \cite{Newman2005,Toda2017}.

Scaling relationships do not only apply to the distribution of a single variable, the population size, but also encompass the way other urban and socio-economic variables respond to changes in size \cite{batty2023}. 
An example of this can be seen in the connection between the concepts of \emph{increasing returns to scale} and \emph{agglomeration effects} in spatial economics and urban theory. %
Increasing returns are characterized by the super-linear scaling of specific socio-economic variables. Their connection to agglomeration effects can be traced back to the work of economist Alfred Marshall at the end of the 19th century \cite{marshall1890growth}.
Facilitated by larger size and greater population density,
the clustering of economic activity in cities has been associated with the advantages of close proximity between economic agents in the form of access to a larger pool of labor, wider ranges of suppliers in intermediary exchanges, and proximity to local markets, leading to shared infrastructure, and lower transport and communication costs \cite{krugman1996urban}.
The advantages of agglomeration can intensify with the concentration of companies of the same industry, leading to regional or urban specialization, including labor. This specialization can also foster knowledge spillovers \emph{within} an industry, known as Marshall-Arrow-Romer externalities or simply \emph{Marshall's scale economies} \cite{marshall1890growth}, promoting innovation and growth in productivity and employment. 

In addition to specialization, scale in urban agglomeration can also lead to industrial \emph{diversity} (\emph{Jacobs economies}
\cite{jacobs1961life,Jacobs1969}). In larger cities, the concentration of economic activity can create a diverse mix of industries. Communication becomes more extensive with decreasing distances \cite{allen1984managing}. 
Combined, economic diversity and proximity encourage cross-fertilization of ideas and technologies through knowledge spillovers \emph{between} industries \cite{Jacobs1969,glaeser1992growth}. 
This setting can lead to a virtuous process where innovation and productivity increase, attracting more businesses to the city, further diversifying the economy, and stimulating economic resilience, like the ability to recover from economic shocks \cite{martin2015notion}. For example, doubling the size of a city by grouping different industries could increase productivity by 3\,\% to 8\,\% \cite{moomaw1981productivity, tabuchi1986urban, rosenthal2004evidence}.
Urban areas with greater industrial diversity experience higher income growth.
More broadly, evidence of the relationship between urban scaling and increasing returns can be found in productivity \cite{glaeser2010complementarity, lobo2013urban}, employment, and income \cite{wheaton2002urban}.
In turn, cities can benefit from agglomeration economies in the provision of public goods and services, such as transportation, education, and healthcare. Due to their greater population density and scale, larger cities are often able to provide these services more efficiently and at a lower cost per capita than smaller cities or rural areas.

We have seen in recent years a larger trans-disciplinary effort from the scientific community to understand urban scaling properties. Apart from %
work on increasing returns such as agglomeration and diversity effects, particularly since the 1960s in urban theory \cite{jacobs1961life} and spatial economics \cite{Jacobs1969} and strongly supported by data since the 1990s \cite{Glaeser1992,rosenthal2003}, more recent contributions have emerged from geography and physics.
This field has gained ground since the turn of the 21st century, thanks mainly to the increase in computational capacity, which allowed an unprecedented acquisition of urban data. It is in this context that more robust interpretations of these laws emerge, mainly driven by the %
work of Denise Pumain \cite{pumain2006}, Luis Bettencourt, Geoffrey West \cite{bettencourt2007growth,Bettencourt2013} and many others \cite[e.g.]{Ribeiro2023} who, based on extensive empirical evidence, were able to formulate theories to explain how urban metrics scale with population size. 
Today, urban scaling is one of the key areas in the so-called new science of cities \cite{Batty2013,batty2023}. The following sections will address more details about these theories and associated empirical evidence.

\section{Empirics}
\label{sec_empirics}

\begin{figure} %
	\begin{center}
        \includegraphics[width=\textwidth]{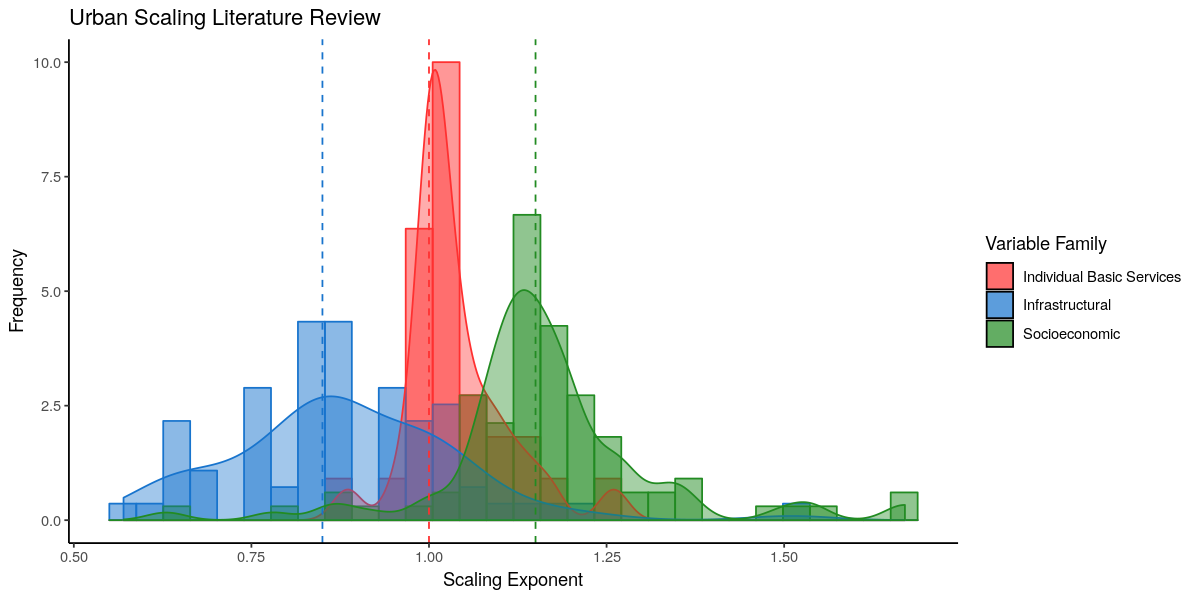}
		\caption{\label{fig_scaling_regime} 
	Distribution of values of the scaling exponent $\beta$ for three different categories/families of urban variables. The samples are composed of data extracted from 70 papers. In total, more than 550 data points were extracted for a plethora of variables. From these data, it is possible to see how variables of each type are distributed in the sub-linear, linear, and super-linear regimes.  The vertical dashed lines represent $\beta = 0.85$ (blue), $=1$ (red), and $=1.15$ (green). }
	\end{center}
\end{figure}

Empirical evidence suggests that the scaling exponent $\beta$ depends strongly on the type/category of the urban variable. 
The seminal work of Bettencourt and colleagues \cite{Bettencourt2007} proposed three categories/families of urban variables related to scaling: (i) socio-economic; (ii) structure and infrastructure; and  (iii) individual needs.
Fig.~(\ref{fig_scaling_regime}) shows the scaling exponent distribution for these three categories when a sample coming from 70 papers is considered
\cite{joao-meta-analysis}.

\begin{figure}[t]
	\sidecaption[t]
	\includegraphics[scale=0.20]{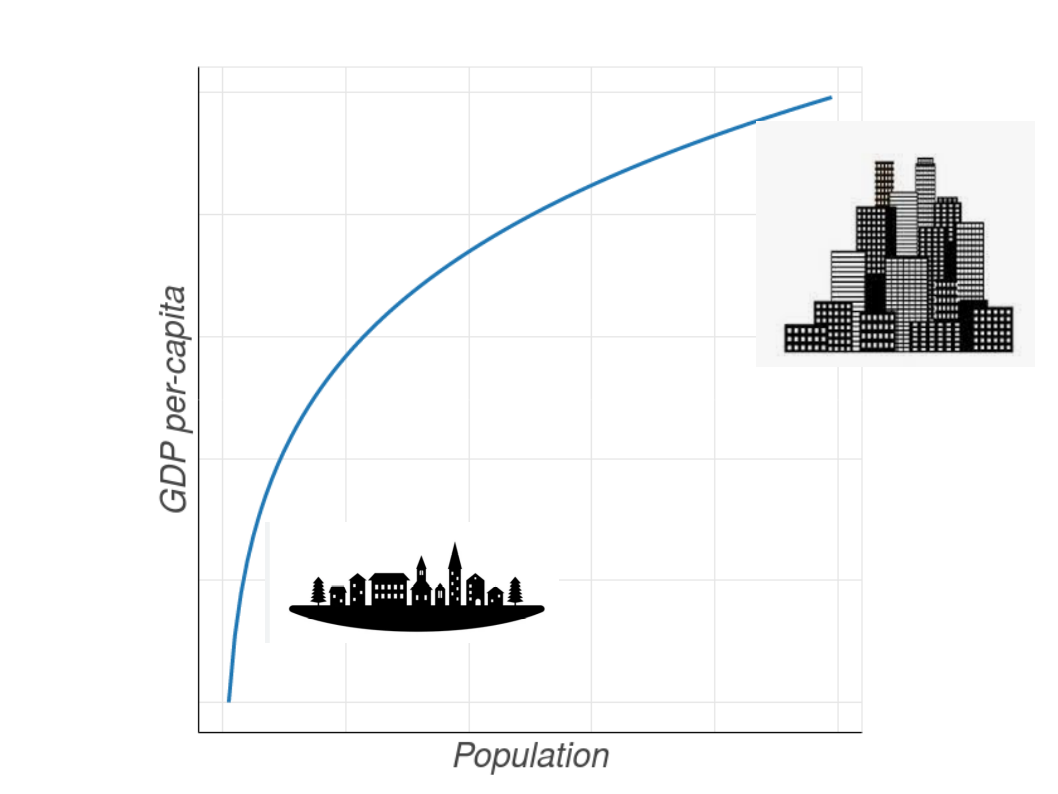}
		\includegraphics[scale=0.20]{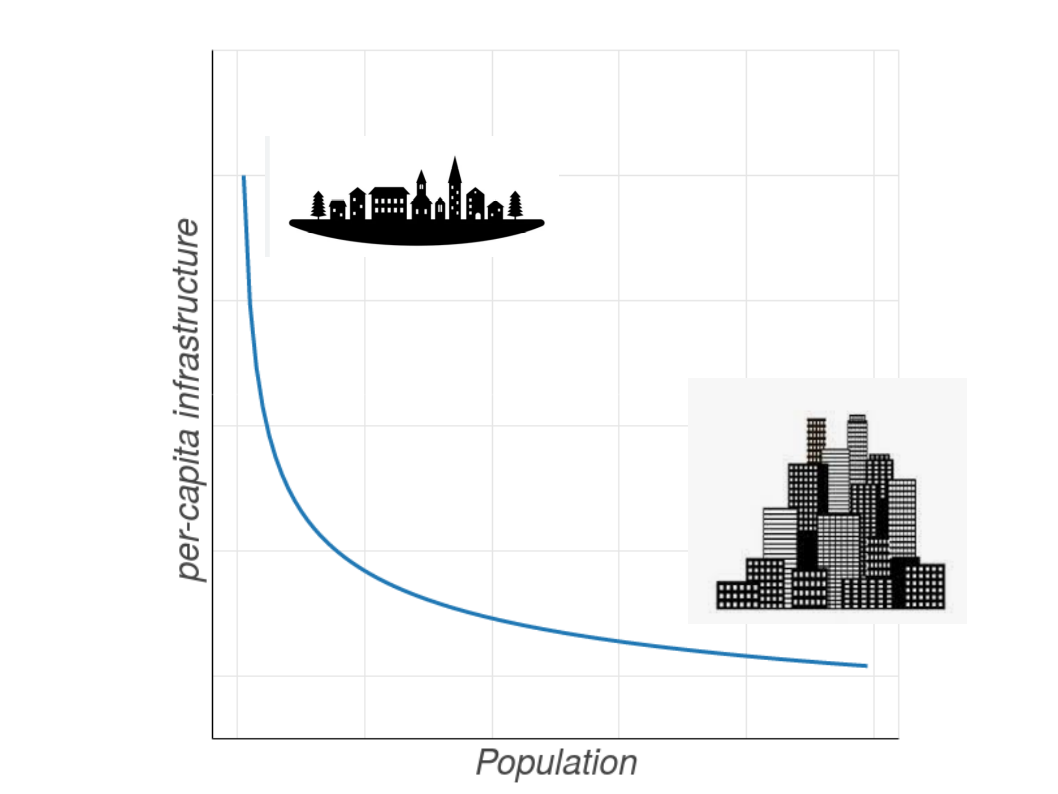}
	\caption{		\label{fig_percapita}
Illustration of per-capita scaling.
Left: typical GDP per capita ($Y/N$) as a function of population size ($N$). The monotonic increasing curve (due to the super-linear scaling of $Y$ with $N$) implies that larger cities have more GDP per inhabitant. In simple words, larger cities are richer.
Right: typical infrastructure per capita ($Y/N$) as a function of population size ($N$). The monotonic decreasing curve (due to sub-linear scaling of $Y$ with $N$) implies that larger cities require less infrastructure per inhabitant. In simple words, larger cities do more with less.}
\end{figure}

The first category of variables is associated with social variables, such as communicable diseases and economic entities or activities (including GDP and wages), in which the scaling exponent is typically distributed around $\beta \approx 1.15$. The fact that this exponent is greater than 1 (the super-linear regime) means that when city size increases, the per-capita number of such variables also increases, as shown in Fig.~\ref{fig_percapita}-left. In simple words, \emph{larger cities are richer}.
The second category of variables is associated with urban structures, such as the total length of streets and total area covered by streets %
(fundamental allometry), and with infrastructures such as the total length of electrical cables or the number of gas stations, 
in which the scaling exponent is distributed typically around $\beta \approx 0.85$. 
The sub-linear scaling exponent means that when city size increases, the per-capita number of these structural and infrastructural %
quantities decreases, as shown in Fig.~\ref{fig_percapita}-right. This implies a scaling economy, which means that \emph{larger cities do more with less}.
A third category of variables is associated with individual needs, such as water and electricity consumption, the number of jobs, and the number of rented houses, 
in which the scaling exponent is distributed typically around $\beta \approx 1$, indicating a linear relationship between these variables and the city population size.
In the following subsections, we present more details about the scaling behavior of each of these categories.

\subsection*{Socio-economic Variables}
The finding $\beta > 1$ for socio-economic variables means that when we increase the size of a city, the GDP or number of interactions increases more than proportionally. 
That is the so-called \emph{increasing returns to scale} in larger cities \cite{joao_plosone2018,Bettencourt2013, ribeirocity2017}.
To understand this, consider the following example. 
Suppose that a specific city increases its population by 1\%, 
then if $\beta = 1.15$,  the GDP of the same city will increase by 1.15\%. 
That is because the derivative of Eq.~(\ref{eq_power_law}) yields $dY = Y_0 \beta N^{\beta -1} dN = Y_0 N^{\beta} \beta dN/N$, and consequently $dY/Y = \beta dN/N$. That means if  $\beta \approx 1.15$, the increment in the urban metric $(dY/Y)$ is approximately 1.15 times bigger than the population increment $(dN/N)$. But please note that this numeric example only works for small population increments.

The increasing returns to scale observed can be attributed to the interplay between properties of the urban structure  and 
human interactions - the key variable driving socio-economic development. Empirical evidence suggests that social interactions grow super-linearly with population size, as evidenced by analyses of phone call data and measures of interpersonal contacts \cite{Schlapfer2014,Newman2005,ribeirocity2017}. This increased interactivity is thought to be a catalyst for creativity and productivity, resulting in social and economic benefits \cite{jacobs1969economy}. Therefore, larger cities with larger populations are likely to provide more opportunities for interactions and contacts between people, leading to wealthier and more creative cities \cite{florida2003cities, gordon2011does,Bettencourt2013,Ribeiro2021b}.
Section \ref{sec_explanations} presents more details about this discussion.

However, the increased connectivity between people in larger cities might have 
a downside. Large cities are more susceptible to epidemics and communicable diseases, such as  AIDS \cite{bettencourt2007growth} and  COVID-19 \cite{Stier2020}, which increase super-linearly with the size of a city.

Alternatively to the increasing return to scale, the super-linearity of socio-economic variables can also be a consequence of the imbalance between larger and smaller cities within a specific system of cities (e.g., a country). In other words, super-linearity can also represent a situation in which wealth is disproportionately concentrated in larger cities, as observed in some countries \cite{Strano2016a,HRibeiro2021}.

\subsection*{Structure and infrastructure Variables}

In terms of structure and infrastructure variables, the scaling exponent is approximately $\beta = 0.85$, as depicted in Fig.~\ref{fig_scaling_regime}. Additionally, the scaling relationship between area and population,  known as \textit{fundamental allometry}, varies %
across different countries but consistently remains below 1 \cite{burger2023}, indicating a sub-linear regime.
In analogy to the increasing returns, $\beta = 0.85$ implies that a small increment in city size results in a per-capita infra/structure reduction of around 15\,\%. For example, when the size of a city increases, the number of gas stations per inhabitant tends to decrease (see Fig.~\ref{fig_percapita}-right). The sub-linear scaling of this kind of variable is usually explained as a consequence of social network properties and spatial constraints, acting in a bottom-up generation process 
\cite{ribeirocity2017,Ribeiro2021b}.

Interestingly, the scaling of urban infra/structure has a qualitative analogy with scaling in biology.
While there is a lower per-capita demand for amenities and services in larger cities, there is also a lower energy-per-cell demand in large animals.
West and his collaborators \cite{West1997,West1999,West2004} justify this analogy by arguing that both living organisms and cities are composed of self-similar and fractal resource distribution networks.
In both systems, there is an underlying network structure (for transport, supply, etc.) that mediates the interactions among cells or among individuals~\cite{Bettencourt2013}.

However, the quantitative analysis differs between cities and biology. While in cities, the scaling exponent is around $\beta = 0.85$ - between infrastructure and population size -, in the case of biology, the exponent is around $0.75$ - between energy expended/demanded and animal mass. This means biology is more efficient than cities: %
while a slight increase in city size leads to a 15\,\% reduction of infra/structure per capita, a slight increase in a living organism leads to a reduction of 25\,\% of energy demanded per cell. We could think that biology is more efficient because it has a much longer evolutionary history than cities. %
It would follow that cities would evolve towards a better economy of scale in the coming centuries or
millennia. However, some empirical studies show that the urban scale exponent %
might remain constant over decades \cite{Ribeiro2020,Xu2020a}, and would not be %
actually evolving. In this sense, perhaps the explanation for this quantitative difference between biology and cities is that they are fundamentally different systems despite presenting some qualitative similarities.
For instance, while in biology individuals are constantly eliminated by natural selection, cities are comprised of highly durable structures that, once built, %
tend to last for decades or centuries, being changed at significant human and material cost \cite{alexander1964notes, netto2017social}.

\subsection*{Individual needs  Variables}
The third type of variable scales linearly with population size ($\beta \approx 1$). These include urban variables related to individual needs, such as water consumption, number of jobs, and number of rented houses. The linearity of these variables means that the size of the city does not affect the per-capita values of these quantities. In other words, it does not matter if a person lives in a small town or in a big city; the amount of water each individual consumes, on average, is the same.
Consequently, the total water consumption in the city, as a whole, grows proportionally to the increase in the number of inhabitants. 
The same situation holds for other individual needs variables, as evidenced by many empirical pieces \cite{bettencourt2007growth}.

\subsection*{Deviations from this categorization}

In addition to the success of this categorization, which has opened doors for a new understanding of urban mechanisms and the application of quantitative methods, it is interesting to mention some works that have questioned the limits of this categorization or even proposed alternative classes.

For example, Strano et al. \cite{Strano2016a} demonstrated that low-income European cities exhibit superlinear scaling between \textit{gross metropolitan product} and population, while high-income cities show linear scaling.
Additionally, Meirelles et al. \cite{joao_plosone2018} identified a fourth category of variables associated with structures and infrastructures that depend strongly on top-down mechanisms under planning, such as public investments or political and economic incentives. In this case, structure and infrastructure variables that would be expected to have a sub-linear scaling exponent, according to Bettencourt et al.'s categorization, are found to have a super-linear scaling exponent due to the action of external factors. This is the case, for instance, of sewage treatment systems and health facilities in Brazil.

The lack of a widely adopted definition for city boundaries is another factor that interferes not only with the predicted values of the scaling exponent but also can lead to transitions between different scaling regimes, such as from a super-linear to a sub-linear regime or vice versa \cite{Cottineau2017,arcaute-batty2015}.
More detailed discussion on this topic can be found in section~\ref{sec_critic}, which delves into criticisms of urban scaling.

\subsection*{Universalities}

Recent works have supported the robustness of the quantitative urban scaling exponent estimates (e.g.\ \cite{Ribeiro2020}). It has been observed not only in different countries, %
but also for different historical periods \cite{ortman2014pre}. These pieces of evidence suggest that the robustness of the urban scaling value could be, in fact, the manifestation of some kind of universality. That is, there may be fundamental principles that govern the growth and organization of cities that are independent of specific cultural, historical, and geographic factors. Optimistically thinking, if this universal law is %
true, then more comprehensive theories can be identified and proposed to explain urban phenomena. Understanding these principles and the underlying processes could have relevant implications for urban planning, sustainability, and the development of cities around the world.

However, there is also evidence against the idea of universality. For example, the ones cited in the last subsection, 
among other evidence \cite{Rybski2017a,Muller2017}. 
After all, the identification and validation of such dynamics, whether universal or not, should help decisively in the formulation of urban policies in order to identify opportunities and improve the quality of life of citizens. Therefore, further research is needed to clarify the possible universality of urban scaling and its implications for urban planning and policy-making.

\section{Possible explanations}
\label{sec_explanations}

Why does urban scaling emerge at all? 
This question is at the core of an important debate in the recent literature. While several models attempt to explain this emergence, a recent review \cite{Ribeiro2023} highlights the lack of consensus on the fundamental process behind the phenomenon of urban scaling. While some models have succeeded in explaining specific aspects of cities, a more fundamental and universal explanation, that is, a \textit{ urban scaling theory},  has yet to be developed. So far, we have a collection of ideas and insights that, at best, explain the scaling of specific urban sectors.

Regarding the fundamental allometry, which defines the sub-linear scaling relationship between cities' area and population, the Bettencourt model \cite{Bettencourt2013} offers an explanation. It posits that individual socioeconomic production should be sufficient to cover transportation costs, which guarantees accessibility and ensures full integration of the city.
This idea conducts that the fundamental alometry scaling exponent, say $\beta_{FA}$, scales sublinearly as $\beta_{FA} = D_f/(D_f +1)$, where $D_f$ is the fractal dimension of the city's building-up area.  
Another model that addresses fundamental allometry is the one proposed by Louf and Barthelemy \cite{Louf2013,louf2014congestion}. They propose that congestion resilience might contribute to the sublinearity between area and population size. 
This idea yields results similar to the Bettencourt model despite having different underlying premises. Further details on this discussion can be found in \cite{Ribeiro2023}.

The underlying principle of urban structures and infrastructures scaling usually stems from network properties. For instance, Bettencourt explains its sublinearity by invoking hierarchical network properties, akin to West, Brown, and Enquist's scaling theory for biological systems (see section~(\ref{sec_hist})).
Molinero \& Thurner  \cite{Molinero2021} add that geometrical considerations are a primary factor influencing urban scaling. They argue that the infra/structure scaling exponent, namely  $\beta_{\rm{infra}}$,  arises from the interplay between two fractal structures: one formed by the spatial distribution of people, characterized by fractal dimension $D_P$, and the other formed by the street network, characterized by fractal dimension \( D_{\rm{infra}} \).
They show this exponent is determined by the ratio between the fractal dimensions of these two structures, that is 
$\beta_{\rm{infra}} = D_{\rm{infra}}/D_P$.  
Since \( D_P \geq D_{\rm{infra}} \) (due to people being embedded in 3D space while streets are in 2D space), the emergence of sub-linearity occurs due to the difference in fractal dimensions of these structures.
Alternatively, Ribeiro et al. \cite{ribeirocity2017} concentrate on the number of amenities required in a city to fulfil the needs of its inhabitants. 
They propose a gravity interaction model between consumers and firms based on distance, characterized by a decay exponent $\gamma$. 
This approach was motivated by the tendency of citizens to choose nearby locations for purchasing goods, resulting in $\beta_{\rm{infra}} = \gamma/D_P$.
This means that urban scaling could be explained by the relation between the accessibility (given by the decay exponent $\gamma$) and the geometric/spatial population distribution properties (characterized by $D_P$).
As a smaller value of $\gamma$ indicates a greater range that people are willing to travel, this result suggests that cities with better mobility promote larger infrastructure scale economies.

\begin{figure}[t]
	\sidecaption[t]
    \includegraphics[width=\textwidth]{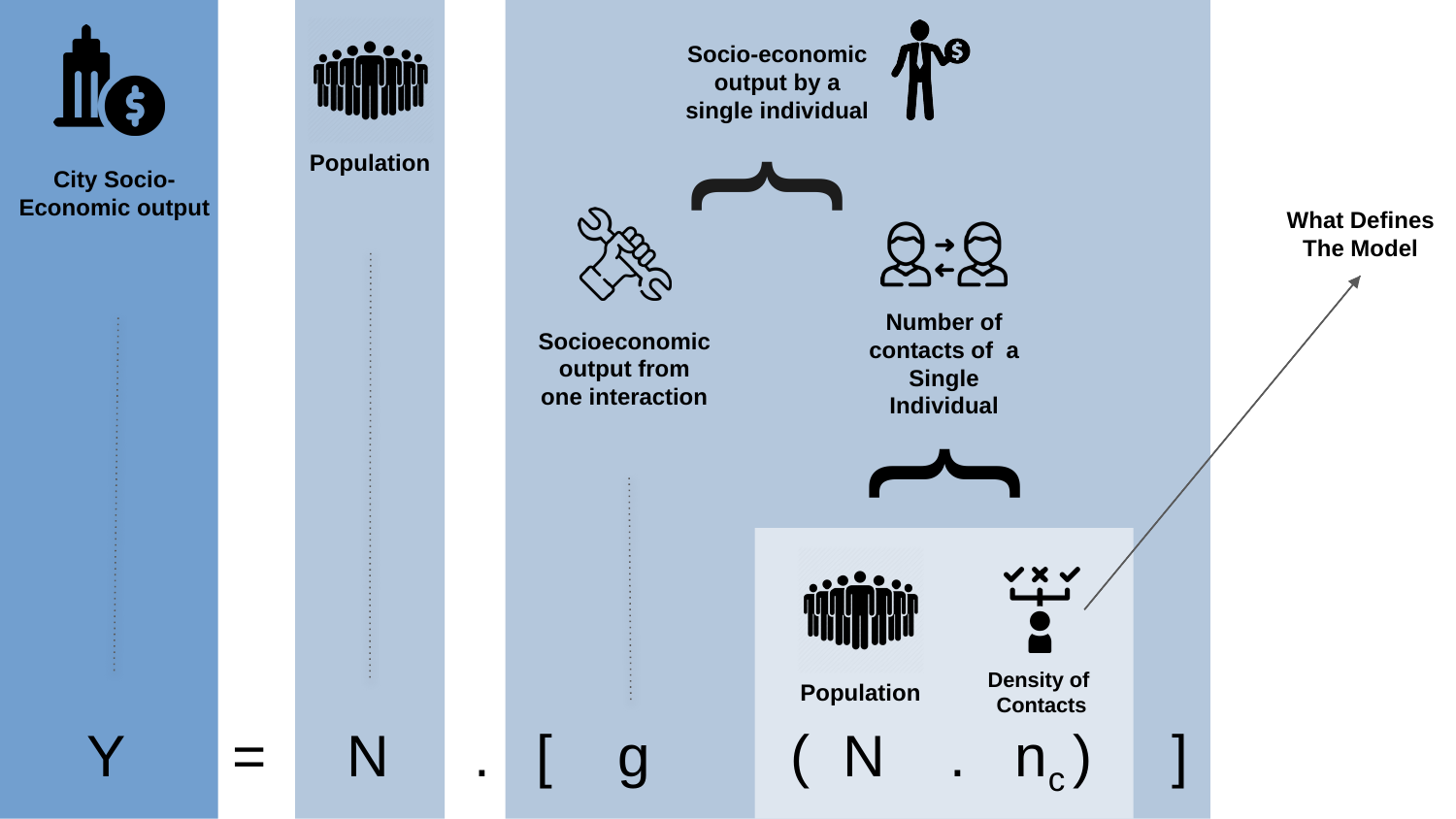}
			\caption{		\label{diagrama_Y}
   Diagram illustrating the general framework for understanding how the number of contacts relates to the socio-economic output $Y$ (e.g., GDP).  
The idea is that   $Y$  can be understood as the sum of all individual socio-economic outputs of the city, that is, $Y = Ny$, where  $y$ is the (average) socioeconomic output by a single individual. In turn, $y$ is the result of the socioeconomic output generated by the interaction between a typical individual and his/her contacts. That is,  $y$ is given by the average number of contacts multiplied by the socio-economic output of the interaction with one single contact; that is,  $y = \langle k_i \rangle g$.
This idea yields the general relation $Y =  g N^2 n_c$, where $n_c = \langle k_i \rangle / N$, is the density of contacts. In fact, the density of contacts $n_c$  is what defines some particular model in the literature. 
Source: \cite[Fig.4]{Ribeiro2023}.
}
\end{figure}

The diversity of ideas presented in the last paragraph 
 makes it clear that there is no consensus about the primary mechanisms that lead to fundamental alometry and infrastructure scaling. 
  However, within the context of socioeconomic variables, it is nearly unanimous in the scientific community that the driving force behind the emergence of super-linear scaling is the \textit{intensity and frequency of human interactions}.
According to empirical findings, the number of contacts people have -- for instance, measured by the number of mobile phone contacts, which grows super-linearly with the city population size \cite{Schlapfer2014}. 
Then, %
hypothetically, a series of other urban metrics, such as GDP, number of patents or cases of socially transmissible diseases, also scale super-linearly. %
The larger the city, the greater the density and, consequently, the more opportunities for people to interact. With a larger number of interactions, more ideas and wealth are created, in the sense that connectivity is the motor that generates increasing returns to scaling \cite{jacobs1969economy, florida2003cities}. Within this context, it is proposed in \cite{Ribeiro2023} a general framework to connect the number of interactions and socio-economic urban scaling, as summarized in the scheme presented in 
Fig.~\ref{diagrama_Y}). 
This scheme yields a general equation 

\begin{equation}\label{eq_general}
Y \sim  N^2 n_c \, ,    
\end{equation}
when it is assumed that 
the socioeconomic output from one interaction, namely $g$, is scale-independent. Here, $n_c$ is the density of contacts, that is,
the ratio between the average number of contacts  (friends, clients, etc..) of a single person, namely $\langle k_i \rangle $, and the city population size; that is $n_c = \langle k_i \rangle / N$.  
In \cite{Ribeiro2023}, it is shown that the general equation~(\ref{eq_general}) embraces many mathematical models in the literature,  and these models differentiate each other only in the way they propose to estimate or determine the density of contacts $n_c$.

For instance, the Bettencourt model \cite{Bettencourt2013} employs cross-sectional ideas to infer human interaction. This model posits that the density of contacts is given by the ratio between the area that a single person effectively uses, say $a$, and the total infrastructured area of the city, say  $A_n$. More especifically, $n_c = a/ A_n$. 
Then if $a$ is scale-invariant  and $A_n \sim N^{\beta_{\rm{infra}}}$ (scaling of the infrastructure, as explained before),  then $Y \sim N^{2-\beta_{\rm{infra}}}$, and consequently the socio-economic scaling exponent, namely $\beta_{\rm{se}}$, is determined by
$\beta_{\rm{se}} = 2- \beta_{\rm{infra}}$.

Another form to infer the density of contacts is to use the gravity idea, that is, considering that the probability of two-person interactions decays with the distance as $p(r) \sim 1/r^\gamma$, as suggested in 
\cite{ribeirocity2017}. The decay exponent $\gamma$ controls the interaction range (as mentioned before). 
What emerges from this model is that the density of contacts scales as $n_c \sim N ^{-\gamma/D_P}$ and consequently, from the general equation~(\ref{eq_general}), one has 
$Y \sim N^{2- \gamma/D_p}$. That is, this idea leads to $\beta_{\rm{se}} = 2- \gamma/D_P$, which means that, once again,  the gravity model explains
urban scaling by the relation between the accessibility and population spatial distribution. 
Note that the superlinearity ($\beta_{\rm{se}}> 1$) only happens if $\gamma$ is sufficiently small (in fact, smaller than $D_P$), which implies that the
superlinearity is a consequency of the 
range of interaction of the people is  sufficiently large.
That is, increase return to scaling is a consequence of a full integrated  city.

What is remarkable is that, despite the Bettencourt and gravity models being fundamentally different, both suggest that the integrity of the city fosters increasing returns to scaling and infrastructure economies. The convergence of different models to the same conclusion, namely the importance of city integrity, implies that the integration of diverse individuals across distances may be a fundamental component of the agglomeration effect.

Indeed, beyond the concepts discussed so far, alternative explanations for urban scaling exist.
In fact, as identified in %
\cite{Ribeiro2023}, there are two groups of models to elucidate this phenomenon: intra- and inter-city models. In intra-city models, the prevailing notion is that non-linearity primarily arises as a consequence of endogenous processes within cities. The examples cited above illustrate this type of model. However, it seems odd to attribute the entire responsibility for inferring socio-economic output solely to internal properties of the city population, given that cities are constantly exchanging goods, services, and people with each other. In this sense, some authors propose models - the intra-city models - that consider exogenous information to explain urban scaling. In these models, factors such as commuting  \cite{Spadon2019,Alves2021}, the influence of nearby cities  \cite{Altmann2020}, and the hierarchical organization expressed by Zipf's law  \cite{HRibeiro2021,Gomez-Lievano2012}, as well as the migration of creative individuals attracted to larger urban centers \cite{Florida2007,Keuschnigg2019a}, are taken into account.

Of course, works in urban theory have dealt with aspects of people's actions and performance in connection with urban structures since Alexander's pioneering works in the 1960s \cite{alexander1964notes, Alexander1966}, a tradition unfolded into approaches to street networks in Hillier's \cite{Hillier1996} space syntax, paralleled by efforts in urban morphology stemming from Jacobs' urban vitality theory \cite{jacobs1961life} and Martin and March's \cite{martin1972urban} work on built form, and by spatial interaction modeling by Batty \cite{batty1976} and others. Indeed, the spatial dimension of cities as a factor with explanatory power over social processes lies at the heart of urban theory. However, the explicit connection between those spatially rich structural views of cities with scaling effects is yet to be fully pursued.

In summary, urban scaling is a complex phenomenon resulting from the %
interplay of factors such as population, density, hierarchical organization, and geometry (expressed by network structure, spatial distributions, %
topography, physical barriers, and fractality) that can either enhance or hinder interactions between people. 
From a more simplistic perspective, we can say that the spatial structure of a city plays a crucial role in determining the nature and extent of human interactions, which in turn shapes the scaling of various urban metrics. However, there is still much more to be understood.

\section{Criticism}
\label{sec_critic}
While the empirical findings described in the preceding sections support the notion that urban metrics are related non-linearly to city population size, there are still counterexamples or contradictions that require further investigation and resolution.

An important issue of urban scaling, as expressed by the power-law given in Eq.~(\ref{eq_power_law}), is that it implicitly assumes that urban metrics are dominated by city population size and do not account for the relationships between cities. Cities are complex open systems, governed by a virtually infinite number of economic, social, and physical processes.
From an optimistic point of view, one could argue that the power-law Eq.~(\ref{eq_power_law}) is a successful first-order approximation and that the interference of exogenous processes is a second-order negligible perturbation. Under this perspective, one could contend that all city complexities vanishes, leaving only one piece of information to infer the urban metric measure, i.e.\ the city population size $N$.
However, from a critical point of view, it is challenging to accept that a city's performance and economy would depend solely on a single variable in an increasingly interconnected contemporary world.%

Firstly, the very nature of this variable, i.e.\ the idea of using population size as a proxy for city size, can be seen as a problem (see \cite{arcaute-batty2015, Rybski2019, cottineau2015paradoxical,Molinero2021}). The representation of cities as zero-dimensional objects -- the absence of the spatial dimension of cities -- in data-driven research ignores two facts: that there are infinite shapes and configurations to accommodate the same population; and that such configurations or structures matter. 
The performance of cities can vary significantly based on their internal structures, even if they share the same population and geographic size.

A step further, cities are social products. Like other complex social phenomena, they resist calls to universality and other possible temptations emanating from the field of social physics. Cities are highly subject to cultural and regional variations that enormously enrich them as artifacts, and urban life as collective experiences. So far, the urban scaling approach has been mostly indifferent to such diversity. Of course, one could argue that this is precisely the point.
However, social and spatial differences %
may have causal effects. %
For instance, different urban structures and the cultural spatial information encoded in them \cite{Netto2023} might help explain why populations in certain world regions achieve larger economic outputs, illustrated in Fig.~\ref{fig_pib}, or are better at spatial navigation \cite{coutrot2022entropy}.

A recurrent and more pragmatic critical point is the dependence of the estimated scaling exponent value on the definition of a city \cite{Arcaute2022,Rybski2019}. Although we can easily identify a city by sight, it is not clear where its boundaries begin and end. Recent works have demonstrated that the definition of the urban boundary affects scaling exponents, and in some cases, their values 
can change the regime/classification that a given metric below  \cite{Cottineau2017,arcaute-batty2015}. As Cottineau et al.\ \cite{Cottineau2017} reported, 
``different scaling regimes can be encountered for the same territory, time, and attribute, depending on the criteria used to delineate cities''.
A striking example of this inconsistency is the analysis of carbon emissions \cite{Louf2014a}. When \emph{urban areas} are used, a super-linear regime is obtained, indicating that \emph{small cities are greener}. However, for \emph{combined statistical areas}, a sub-linear regime is obtained, indicating the opposite, that \emph{larger cities are greener}. 
Besides the city boundaries definition, the selection of statistical methods used in the analysis can also impact the estimate of the scaling exponent, as highlighted in a study by Leitão et al.\ \cite{leitao2016}. Therefore, further research is needed to shed light on this aspect.
For a more complete discussion of the deficiencies and criticisms of urban scaling laws, %
see \cite{Arcaute2022}.

In short, it is fair to say that there are many factors beyond population size that can promote gains in socio-economic performance or optimization in urban structures and infrastructures, as seen in decades of research in urban theory (e.g.\ \cite{martin1972urban, Hillier1996, Batty2013}), and in the increasing returns literature in spatial economics (e.g.\ \cite{marshall1890growth, jacobs1969economy, glaeser1992growth}, to mention a few). %

\section{Open issues/questions}
\label{sec_opnequestions}
The apparent universality in some aspects of urban phenomena, particularly with regards to the scaling hypothesis discussed here, paves the way for the creation of a \emph{Unified Urban Theory} (UUT) \cite{Ribeirofisica2020, Bettencourt2010,Lobo2020}. The UUT would be a quantitative theory that combines a few premises and derives many different observed urban patterns as particular cases \cite{hypotheses2021}. The possibility of achieving such a feat indicates that, with the advent of big data and high computational performance, urban science is gaining the status of an ``quantitative science''.
If  UUT can indeed be developed, it would provide a set of systematic approaches that can be used to explain and predict urban features, benefiting planners and public administrators. As we argued above, the crucial challenge here lies in the very nature of cities. Like other social phenomena, cities have substantial cultural differences, both physical and symbolic, full of causal potential, which easily evade general theories.

Another issue is whether scaling laws in urban phenomena are analogous to those in biological systems. For instance, are the scaling laws in biology and urban scaling laws manifestations of the same physical principle? If so, what principle is this? If not, what differentiates these two systems? And how could we demonstrate it (or not) in a technical and quantified manner without relying on mere speculation? Questions like these may well guide future investigations.
For instance, scaling in biology appears to be analogous to urban structure and infrastructure scaling, as both present sub-linear scaling. However, their scaling exponent values differ quantitatively. While data and theory on vascular organisms suggest that $\beta \approx 0.75$ (see Sections~(\ref{sec_intro}) and~(\ref{sec_hist})), in the case of urban infrastructure, we have $\beta \approx 0.85$.
This quantitative difference implies that biology is more efficient in terms of resource economy. But why is there such a quantitative difference? One naive answer could be that biology had billions of years to evolve while cities only had a few thousand years. Based on this statement, one could argue that cities still need to evolve in order to be as scaling-efficient as biology. However, the data also suggest some kind of stability of the urban scaling exponent when the time evolution of urban systems is analyzed, suggesting that urban systems are already at the equilibrium scaling stage. This shows that the analogy between biology and cities is still a very open point that must be elucidated.

Finally, in response to a crucial critical issues raised above, we also think that the question of scaling properties and effects must be connected with the fundamental problem of how spatial structures become part of the socio-economic dynamics cities express and support. A key question here is how such structures might interfere with the role of population size in harnessing or hindering its power for super-linear or sub-linear behavior as positive effects on socio-economic performance and sustainability. Urban structures characterized by poor distributions of access and opportunities make people’s movement and efforts more difficult, fragile, and harmful to society and its environment, e.g.\ being more subject to fluctuations like traffic congestion and unpredictability such as natural or man-made disasters. Cities are vital in that sense: sustainable societies need well-structured cities.

\section{Summary}
\label{sec_summ}
We discussed in this chapter recent findings, ongoing advancements, and open questions on urban scaling. Exploring parallels to other research areas in urban studies and spatial economics, our overview covered the following key points.
\\
\begin{itemize}
\item %
Following the development of scaling law concepts in biology, there has been a recent trans-disciplinary effort to understand urban scaling properties. Contributions from geography, physics, and computational analysis have led to the formulation of theories explaining how urban metrics scale with population size. However, scaling relationships extend beyond population size and encompass other urban and socio-economic variables. Concepts like increasing returns to scale and agglomeration effects in spatial economics  %
and urban theory have captured how scale, along with proximity and diversity, contribute to productivity, innovation, and economic growth.

\item %
Different types of urban variables have consistently shown specific behaviors regarding scaling and can be grouped into three categories: (i) socio-economic variables (super-linear); (ii) urban structures and infrastructures (sub-linear); and (iii) individual needs (linear).

\item The consistency of urban scaling exponent estimates across different countries and historical periods suggests the possibility of a universal phenomenon governing the growth and organization of cities, independent of cultural, historical, and geographic factors. However, some evidence also suggests that cultural and political conditions may challenge universality in urban scaling. More research is necessary to determine whether urban scaling is influenced by specific contextual factors. 

\item  %
The number of human interactions is believed to be the driving force behind super-linear scaling of socio-economic variables. However, there is little consensus on the fundamental processes behind the details.
Natural or artificial features, such as efficient transport infrastructure or geographic factors, might interfere with human interaction and impact scaling effects. The spatial structure of a city may also play a crucial role in shaping human interactions and, consequently, the scaling of urban metrics. As a complex phenomenon, urban scaling can be influenced by population, densities, urban structures, hierarchical organization -- and possibly more.

\item  %
Critical observations pointed out that the definition of a city's boundaries has a significant impact on scaling exponent estimates. Furthermore, relying solely on a single variable, i.e.\ population size, to explain a city's performance and economy is challenging. Population size as a proxy for city size overlooks the spatial dimension of cities and the diverse internal structures they can have, which can significantly impact interactions. %

\item  Open questions and future directions include the possibility of creating a unified theory, which is particularly challenging when context-dependent features that help defining the unique identities of cities as social products are considered. That said, the similarities and differences between scaling laws in biology and urban phenomena continue to appeal as key areas of investigation, including the time evolution of urban systems and the stability of the scaling exponent. And a major research avenue begs the question of how durable urban structures may harness or hinder the power of population size in super-linear or sub-linear effects on socio-economic and environmental sustainability. 

\end{itemize}



\begin{thebibliography}{10}
	
	\bibitem{Kadanoff2000}
	L.~P. Kadanoff, {\em {Statistical physics: statics, dynamics and
			remormalization.}}
	\newblock Singapore: World Scientific, 2000.
	
	\bibitem{Stanley1987}
	H.~E. Stanley, {\em {Introduction to Phase Transitions and Critical
			Phenomena}}.
	\newblock Oxford: Oxford University Press, 1987.
	
	\bibitem{West2017}
	G.~West, {\em {Scale: The Universal Laws of Growth, Innovation, Sustainability,
			and the Pace of Life in Organisms, Cities, Economies, and Companies}}.
	\newblock 2017.
	
	\bibitem{Allman1999}
	J.~Allman, {\em {Evolving Brains}}.
	\newblock 1999.
	
	\bibitem{Hemmingsen1960}
	A.~Hemmingsen, ``{Energy metabolism as related to body size and respiratory
		surfaces, and its evolution.},'' {\em Rep Steno Mem Hosp Nord Insulinlab},
	vol.~9, no.~1-110, 1960.
	
	\bibitem{Ribeiro2022}
	F.~L. Ribeiro and W.~R. Pereira, ``{a Gentle Introduction To Scaling Relations
		in Biological Systems},'' {\em Revista Brasileira de Ensino de Fisica},
	vol.~44, 2022.
	
	\bibitem{West1997}
	B.~J.~E. {Geoffrey B. West, James H. Brown}, ``{A General Model for the Origin
		of Allometric Scaling Laws in Biology},'' {\em Science}, vol.~276,
	pp.~122--126, apr 1997.
	
	\bibitem{shiftspnas2010}
	J.~P. Delong, J.~G. Okie, M.~E. Moses, R.~M. Sibly, and J.~H. Brown, ``{Shifts
		in metabolic scaling , production , and ef fi ciency across major
		evolutionary transitions of life},'' no.~14, 2010.
	
	\bibitem{bettencourt2007growth}
	L.~M.~A. Bettencourt, J.~Lobo, D.~Helbing, C.~K{\"{u}}hnert, and G.~B. West,
	``{Growth, innovation, scaling, and the pace of life in cities.},'' {\em
		Proceedings of the National Academy of Sciences of the United States of
		America}, vol.~104, pp.~7301--6, apr 2007.
	
	\bibitem{Bettencourt2013}
	L.~M.~A. Bettencourt, ``{The origins of scaling in cities},'' {\em Science},
	vol.~340, no.~6139, pp.~1438--41, 2013.
	
	\bibitem{ribeirocity2017}
	F.~L. Ribeiro, {Joao Meirelles}, F.~F. Ferreira, and C.~R. Neto, ``{A model of
		urban scaling laws based on distance-dependent interactions},'' {\em Royal
		Society Open Science}, vol.~4, no.~160926, 2017.
	
	\bibitem{Ribeiro2023}
	F.~L. Ribeiro and D.~Rybski, ``{Mathematical models to explain the origin of
		urban scaling laws},'' {\em Physics Reports}, vol.~1012, pp.~1--39, 2023.
	
	\bibitem{Newman2005}
	M.~Newman, ``{Power laws, Pareto distributions and Zipf's law},'' {\em
		Contemporary physics}, no.~1, 2005.
	
	\bibitem{barabasi-book}
	A.-L. BARAB{\'{A}}SI, {\em {NETWORK SCIENCE THE SCALE-FREE PROPERTY}}.
	
	\bibitem{west-new-york}
	W.~Does, N.~York, H.~So, and M.~Doctors, ``{Why New York Is Just an Average
		City},'' no.~50, pp.~1--10, 2017.
	
	\bibitem{Robiquet1839}
	T.~Robiquet, ``{Rapport sur un m{\'{e}}moire address{\'{e}} al?Acad{\'{e}}mie
		Royale de M{\'{e}}decin par MM Sarrus et Rameaux.},'' {\em Bull Acad R Med
		Belg}, vol.~3, pp.~1094--1100, 1839.
	
	\bibitem{Rubner1883}
	M.~Rubner, 
	{\em Z. Biol.}, vol.~19, pp.~536--562, 1883.
	
	\bibitem{bertalanffy1957}
	L.~von Bertalanffy, ``{Quantitative Laws in Metabolism and Growth},'' {\em The
		Quarterly Review of Biology}, vol.~32, no.~3, pp.~217--231, 1949.
	
	\bibitem{Krogh1916}
	A.~T. Krogh, {\em {Respiratory Exchange of Animals and Man.}}
	\newblock London, UK,: Longmans, 1916.
	
	\bibitem{Kleiber1932}
	M.~Kleiber, ``{Body size and metabolism},'' {\em Hilgardia}, vol.~6,
	pp.~315--353, jan 1932.
	
	\bibitem{DeLong2010}
	J.~P. DeLong, J.~G. Okie, M.~E. Moses, R.~M. Sibly, and J.~H. Brown, ``{Shifts
		in metabolic scaling, production, and efficiency across major evolutionary
		transitions of life},'' {\em Proceedings of the National Academy of Sciences
		of the United States of America}, vol.~107, no.~29, pp.~12941--12945, 2010.
	
	\bibitem{West1999}
	G.~B. West, ``{The Fourth Dimension of Life: Fractal Geometry and Allometric
		Scaling of Organisms},'' {\em Science}, vol.~284, pp.~1677--1679, jun 1999.
	
	\bibitem{West2004}
	G.~B. West and J.~H. Brown, ``{Life's Universal Scaling Laws},'' {\em Physics
		Today}, no.~September, 2004.
	
	\bibitem{Savage2008}
	V.~M. Savage, E.~J. Deeds, and W.~Fontana, ``{Sizing up allometric scaling
		theory},'' {\em PLoS Computational Biology}, vol.~4, no.~9, 2008.
	
	\bibitem{zipf1932}
	{G. Kingsley Zipf}, {\em {Selected studies of the principle of relative
			frequency in language}}.
	\newblock Harvard university press, 1932.
	
	\bibitem{zipf-book-1949}
	G.~K. Zipf, {\em {Human Behavior and the Principle of Least Effort: An
			Introduction to Human Ecology}}.
	\newblock 1949.
	
	\bibitem{Auerbach1913}
	F.~Auerbach and A.~Ciccone, ``{The Law of Population Concentration},'' {\em
		Environment and Planning B: Urban Analytics and City Science}, vol.~50,
	no.~2, pp.~290--298, 2023.
	
	\bibitem{diego2023}
	D.~Rybski and A.~Ciccone, ``{Auerbach , Lotka , Zipf – pioneers of power-law
		city-size distributions},'' pp.~1--7, 2023.
	
	\bibitem{Toda2017}
	A.~A. Toda, X.~Gabaix, S.~Gilchrist, R.~Kirpalani, and N.~Kocherlakota, ``{Zipf
		' s Law : A Microfoundation},'' no.~2011, pp.~1--47, 2017.
	
	\bibitem{batty2023}
	M.~Batty, ``{Scaling in city size distributions},'' {\em Environment and
		Planning B: Urban Analytics and City Science}, vol.~0, no.~0, pp.~1--3, 2023.
	
	\bibitem{marshall1890growth}
	A.~MARSHALL, ``{The Growth of Economic Science, Appendix B of the Principles of
		Economics},'' 1890.
	
	\bibitem{krugman1996urban}
	P.~Krugman, ``Urban concentration: the role of increasing returns and transport
	costs,'' {\em International Regional Science Review}, vol.~19, no.~1-2,
	pp.~5--30, 1996.
	
	\bibitem{jacobs1961life}
	J.~Jacobs, {\em {The life and death of great American cities}}.
	\newblock New York: Random House, 1961.
	
	\bibitem{Jacobs1969}
	J.~Jacobs, {\em {The economy of cities}}.
	\newblock New York: Random house, 1969.
	
	\bibitem{allen1984managing}
	T.~J. Allen {\em et~al.}, ``Managing the flow of technology: Technology
	transfer and the dissemination of technological information within the r\&d
	organization,'' {\em MIT Press Books}, vol.~1, 1984.
	
	\bibitem{glaeser1992growth}
	E.~L. Glaeser, H.~D. Kallal, J.~A. Scheinkman, and A.~Shleifer, ``Growth in
	cities,'' {\em Journal of political economy}, vol.~100, no.~6,
	pp.~1126--1152, 1992.
	
	\bibitem{martin2015notion}
	R.~Martin and P.~Sunley, ``On the notion of regional economic resilience:
	conceptualization and explanation,'' {\em Journal of economic geography},
	vol.~15, no.~1, pp.~1--42, 2015.
	
	\bibitem{moomaw1981productivity}
	R.~L. Moomaw, ``Productivity and city size: a critique of the evidence,'' {\em
		The Quarterly Journal of Economics}, vol.~96, no.~4, pp.~675--688, 1981.
	
	\bibitem{tabuchi1986urban}
	T.~Tabuchi, ``Urban agglomeration, capital augmenting technology, and labor
	market equilibrium,'' {\em Journal of Urban Economics}, vol.~20, no.~2,
	pp.~211--228, 1986.
	
	\bibitem{rosenthal2004evidence}
	S.~S. Rosenthal and W.~C. Strange, ``Evidence on the nature and sources of
	agglomeration economies,'' in {\em Handbook of regional and urban economics},
	vol.~4, pp.~2119--2171, Elsevier, 2004.
	
	\bibitem{glaeser2010complementarity}
	E.~L. Glaeser and M.~G. Resseger, ``The complementarity between cities and
	skills,'' {\em Journal of Regional Science}, vol.~50, no.~1, pp.~221--244,
	2010.
	
	\bibitem{lobo2013urban}
	J.~Lobo, L.~M. Bettencourt, D.~Strumsky, and G.~B. West, ``Urban scaling and
	the production function for cities,'' {\em PLoS One}, vol.~8, no.~3,
	p.~e58407, 2013.
	
	\bibitem{wheaton2002urban}
	W.~C. Wheaton and M.~J. Lewis, ``Urban wages and labor market agglomeration,''
	{\em Journal of Urban Economics}, vol.~51, no.~3, pp.~542--562, 2002.
	
	\bibitem{Glaeser1992}
	E.~Glaeser, H.~Kallal, J.~Scheinkman, and A.~Shleifer, ``{Growth in cities},''
	{\em Journal of Political Economy}, no.~100, pp.~1126--1152, 1992.
	
	\bibitem{rosenthal2003}
	S.~S. Rosenthal, ``{Evidence on the Nature and Sources of Agglomeration
		Economies Stuart},'' {\em Handbook of Urban And Regional Economics}, 2003.
	
	\bibitem{pumain2006}
	L.~J. {Pumain D, Paulus F, Vacchiani-Marcuzzo C}, ``{An evolutionary theory for
		interpreting urban scaling laws},'' {\em Cybergeo: European Journal of
		Geography;}, 2006.
	
	\bibitem{Batty2013}
	M.~Batty, {\em {The new science of cities}}.
	\newblock MIT Press,, 2013.
	
	\bibitem{Bettencourt2007}
	L.~M. Bettencourt, J.~Lobo, and D.~Strumsky, ``{Invention in the city:
		Increasing returns to patenting as a scaling function of metropolitan
		size},'' {\em Research Policy}, vol.~36, no.~1, pp.~107--120, 2007.
	
	\bibitem{joao-meta-analysis}
	{Fabiano L. Ribeiro} and {Joao Meirelles}, ``{Meta-analysis os the categories
		of urban variables related to scaling},'' {\em in preparation}, 2024.
	
	\bibitem{joao_plosone2018}
	J.~Meirelles, C.~R. Neto, F.~F. Ferreira, F.~L. Ribeiro, and C.~R. Binder,
	``{Evolution of urban scaling: Evidence from Brazil},'' {\em PLOS ONE},
	vol.~13, p.~e0204574, oct 2018.
	
	\bibitem{Schlapfer2014}
	M.~Schl{\"{a}}pfer, L.~M.~a. Bettencourt, S.~Grauwin, M.~Raschke, R.~Claxton,
	Z.~Smoreda, G.~B. West, and C.~Ratti, ``{The scaling of human interactions
		with city size.},'' {\em Journal of the Royal Society, Interface / the Royal
		Society}, vol.~11, no.~98, pp.~20130789--, 2014.
	
	\bibitem{jacobs1969economy}
	J.~Jacobs, {\em The economy of cities}.
	\newblock Vintage, 1969.
	
	\bibitem{florida2003cities}
	R.~Florida, ``Cities and the creative class,'' {\em City \& community}, vol.~2,
	no.~1, pp.~3--19, 2003.
	
	\bibitem{gordon2011does}
	P.~Gordon and S.~Ikeda, ``Does density matter?,'' in {\em Handbook of creative
		cities}, Edward Elgar Publishing, 2011.
	
	\bibitem{Ribeiro2021b}
	F.~L. Ribeiro and D.~Rybski, ``{Mathematical models to explain the origin of
		urban scaling laws: a synthetic review},'' {\em arXiv preprint}, pp.~1--22,
	2021.
	
	\bibitem{Stier2020}
	A.~J. Stier, M.~G. Berman, and L.~M.~A. Bettencourt, ``{COVID-19 attack rate
		increases with city size},'' {\em arXiv preprint}, 2020.
	
	\bibitem{Strano2016a}
	E.~Strano and V.~Sood, ``{Rich and poor cities in Europe. An urban scaling
		approach to mapping the European economic transition},'' {\em PLoS ONE},
	vol.~11, no.~8, pp.~1--8, 2016.
	
	\bibitem{HRibeiro2021}
	H.~V. Ribeiro, M.~Oehlers, A.~I. Moreno-monroy, P.~Kropp, and D.~Rybski,
	``{Effects of population distribution on urban scaling},'' {\em PLoS ONE},
	vol.~16, no.~1, 2021.
	
	\bibitem{burger2023}
	J.~R. Burger, J.~G. Okie, I.~A. Hatton, V.~P. Weinberger, M.~Shrestha, K.~J.
	Liedtke, A.~S. M.~G. Kibria, K.~C. Ernst, B.~J. Enquist, and D.~Kendal,
	``{Global city densities : Re- examining urban scaling theory},''
	no.~December, pp.~1--10, 2022.
	
	\bibitem{Ribeiro2020}
	F.~L. Ribeiro, J.~Meirelles, V.~M. Netto, C.~R. Neto, and A.~Baronchelli, ``{On
		the relation between Transversal and Longitudinal Scaling in Cities},'' {\em
		PLoS ONE}, pp.~1--20, 2020.
	
	\bibitem{Xu2020a}
	G.~Xu, Z.~Xu, Y.~Gu, W.~Lei, Y.~Pan, J.~Liu, and L.~Jiao, ``{Scaling laws in
		intra-urban systems and over time at the district level in Shanghai,
		China},'' {\em Physica A: Statistical Mechanics and its Applications},
	vol.~560, no.~September, p.~125162, 2020.
	
	\bibitem{alexander1964notes}
	C.~Alexander, {\em Notes on the Synthesis of Form}, vol.~5.
	\newblock Harvard University Press, 1964.
	
	\bibitem{netto2017social}
	V.~M. Netto, {\em The social fabric of cities}, vol.~250.
	\newblock Routledge New York, 2017.
	
	\bibitem{Cottineau2017}
	C.~Cottineau, E.~Hatna, E.~Arcaute, and M.~Batty, ``{Diverse cities or the
		systematic paradox of Urban Scaling Laws},'' {\em Computers, Environment and
		Urban Systems}, vol.~63, pp.~80--94, 2017.
	
	\bibitem{arcaute-batty2015}
	E.~Arcaute, E.~Hatna, P.~Ferguson, H.~Youn, A.~Johansson, and M.~Batty,
	``{Constructing cities, deconstructing scaling laws},'' {\em Journal of The
		Royal Society Interface}, vol.~12, no.~102, p.~20140745, 2015.
	
	\bibitem{ortman2014pre}
	S.~G. Ortman, A.~H.~F. Cabaniss, J.~O. Sturm, and L.~M.~A. Bettencourt, ``{The
		pre-history of urban scaling},'' {\em PloS one}, vol.~9, no.~2, p.~e87902,
	2014.
	
	\bibitem{Rybski2017a}
	D.~Rybski, D.~E. Reusser, A.~L. Winz, C.~Fichtner, T.~Sterzel, and J.~P. Kropp,
	``{Cities as nuclei of sustainability?},'' {\em Environment and Planning B:
		Urban Analytics and City Science}, vol.~44, no.~3, pp.~425--440, 2017.
	
	\bibitem{Muller2017}
	N.~Z. Muller and A.~Jha, ``{Does environmental policy affect scaling laws
		between population and pollution? Evidence from American metropolitan
		areas},'' {\em PloS one}, vol.~12, no.~8, p.~e0181407, 2017.
	
	\bibitem{Louf2013}
	R.~Louf and M.~Barthelemy, ``{Modeling the polycentric transition of cities},''
	{\em Physical Review Letters}, vol.~111, no.~19, 2013.
	
	\bibitem{louf2014congestion}
	R.~Louf and M.~Barthelemy, ``{How congestion shapes cities: from mobility
		patterns to scaling},'' {\em Scientific reports}, vol.~4, 2014.
	
	\bibitem{Molinero2021}
	C.~Molinero and S.~Thurner, ``{How the geometry of cities explains urban
		scaling laws},'' {\em Interface}, vol.~18, 2021.
	
	\bibitem{Spadon2019}
	G.~Spadon, A.~C. de~Carvalho, J.~F. Rodrigues-Jr, and L.~G. Alves,
	``{Reconstructing commuters network using machine learning and urban
		indicators},'' {\em Scientific Reports}, vol.~9, no.~1, pp.~1--13, 2019.
	
	\bibitem{Alves2021}
	L.~G. Alves, D.~Rybski, and H.~V. Ribeiro, ``{Commuting network effect on urban
		wealth scaling},'' {\em Scientific Reports}, vol.~11, no.~1, pp.~1--10, 2021.
	
	\bibitem{Altmann2020}
	E.~G. Altmann, ``{Spatial interactions in urban scaling laws},'' {\em PloS
		one}, vol.~15, no.~12, pp.~1--12, 2020.
	
	\bibitem{Gomez-Lievano2012}
	A.~Gomez-Lievano, H.~J. Youn, and L.~M. Bettencourt, ``{The statistics of urban
		scaling and their connection to Zipf's law},'' {\em PLoS ONE}, vol.~7, no.~7,
	2012.
	
	\bibitem{Florida2007}
	R.~Florida, The Flight of the Creative Class: The New Global Competition
			for Talent. Harper Business, 2007.
	
	\bibitem{Keuschnigg2019a}
	M.~Keuschnigg, S.~Mutgan, and P.~Hedstr{\"{o}}m, ``{Urban scaling and the
		regional divide},'' {\em Science Advances}, vol.~5, no.~1, 2019.
	
	\bibitem{Alexander1966}
	C.~Alexander, ``{A city is not a tree},'' {\em City}, vol.~122, no.~1,
	pp.~58--62, 1966.
	
	\bibitem{Hillier1996}
	B.~Hillier, {\em {Space is the Machine}}.
	\newblock Cambridge: Cambridge University Press., 1996.
	
	\bibitem{martin1972urban}
	L.~Martin and L.~March, {\em Urban space and structures}, vol.~(282).
	\newblock Cambridge University Press, Cambridge, 1972.
	
	\bibitem{batty1976}
	M.~Batty, ``{Urban Modelling},'' 1976.
	
	\bibitem{Rybski2019}
	D.~Rybski, E.~Arcaute, and M.~Batty, ``{Urban scaling laws},'' {\em Environment
		and Planning B: Urban Analytics and City Science}, vol.~46, no.~9,
	pp.~1605--1610, 2019.
	
	\bibitem{cottineau2015paradoxical}
	C.~C. Cottineau, E.~Hatna, E.~Arcaute, and M.~Batty, ``{Paradoxical
		Interpretations of Urban Scaling Laws},'' {\em arXiv preprint
		arXiv:1507.07878}, no.~July, pp.~1--21, 2015.
	
	\bibitem{Netto2023}
	V.~M. Netto, E.~Brigatti, and C.~Cacholas, ``{From urban form to information:
		Cellular configurations in different spatial cultures},'' {\em Environment
		and Planning B: Urban Analytics and City Science}, vol.~50, no.~1,
	pp.~146--161, 2023.
	
	\bibitem{coutrot2022entropy}
	A.~Coutrot, E.~Manley, S.~Goodroe, C.~Gahnstrom, G.~Filomena, D.~Yesiltepe,
	R.~C. Dalton, J.~M. Wiener, C.~H{\"o}lscher, M.~Hornberger, {\em et~al.},
	``Entropy of city street networks linked to future spatial navigation
	ability,'' {\em Nature}, vol.~604, no.~7904, pp.~104--110, 2022.
	
	\bibitem{Arcaute2022}
	E.~Arcaute and J.~J. Ramasco, ``{Recent advances in urban system science:
		Models and data},'' {\em PLoS ONE}, vol.~17, no.~8 August, pp.~1--16, 2022.
	
	\bibitem{Louf2014a}
	R.~Louf and M.~Barthelemy, ``{Scaling: lost in the smog},'' {\em Environment
		and Planning B: Planning and Design}, vol.~41, no.~5, pp.~767--769, 2014.
	
	\bibitem{leitao2016}
	J.~C. Leit{\~{a}}o, J.~M. Miotto, M.~Gerlach, and E.~G. Altmann, ``{Is this
		scaling nonlinear?},'' {\em Royal Society Open Science}, vol.~3, 2016.
	
	\bibitem{Ribeirofisica2020}
	F.~L. Ribeiro, ``{F{\'{i}}sica das Cidades},'' {\em Revista de Morfologia
		Urbana}, vol.~8, no.~1, p.~e00159, 2020.
	
	\bibitem{Bettencourt2010}
	L.~Bettencourt and G.~West, ``{A unified theory of urban living.},'' {\em
		Nature}, vol.~467, pp.~912--3, oct 2010.
	
	\bibitem{Lobo2020}
	J.~Lobo, M.~Alberti, M.~Allen-Dumas, E.~Arcaute, M.~Barthelemy, L.~A.
	{Bojorquez Tapia}, S.~Brail, L.~Bettencourt, A.~Beukes, W.~Chen, R.~Florida,
	M.~Gonzalez, N.~Grimm, M.~Hamilton, C.~Kempes, C.~E. Kontokosta,
	C.~Mellander, Z.~P. Neal, S.~Ortman, D.~Pfeiffer, M.~Price, A.~Revi,
	C.~Rozenblat, D.~Rybski, M.~Siemiatycki, S.~T. Shutters, M.~E. Smith, E.~C.
	Stokes, D.~Strumsky, G.~West, D.~White, J.~Wu, V.~C. Yang, A.~York, and
	H.~Youn, ``{Urban Science: Integrated Theory from the First Cities to
		Sustainable Metropolises},'' {\em SSRN Electronic Journal}, 2020.
	
	\bibitem{hypotheses2021}
	F.~L. Ribeiro, J.~Lobo, and D.~Rybski, ``{Zipf's law and urban scaling:
		Hypotheses towards a Unified Urban Theory},'' {\em arXiv preprint}, pp.~1--3,
	2021.
	
\end{thebibliography}

\end{document}